\documentclass[prb,twocolumn,superscriptaddress,floatfix,letterpaper]{revtex4}
\usepackage{graphicx}
\usepackage{amsmath, amsthm}
\pdfoutput=1

\begin{document}

\title{
Chiral spin states in polarized kagome spin systems
with spin-orbit coupling
}

\author{Jia-Wei Mei}
\affiliation{Institute for Advanced Study, Tsinghua
University, Beijing, 100084, P. R. China}
\affiliation{Department of Physics, Massachusetts Institute of Technology, Cambridge, Massachusetts 02139, USA}

\author{Evelyn Tang}
\affiliation{Department of Physics, Massachusetts Institute of Technology, Cambridge, Massachusetts 02139, USA}

\author{Xiao-Gang Wen}
\affiliation{Department of Physics, Massachusetts Institute of Technology, Cambridge, Massachusetts 02139, USA}
\affiliation{Institute for Advanced Study, Tsinghua
University, Beijing, 100084, P. R. China}

\date{Nov, 2010}
\begin{abstract}
We study quantum spin systems with a proper combination of geometric
frustration, spin-orbit coupling and ferromagnetism.  We argue that such a
system is likely to be in a chiral spin state, a fractional quantum Hall (FQH) state for
bosonic spin degrees of freedom.  The energy scale of the bosonic FQH state is
of the same order as the spin-orbit coupling and ferromagnetism --- overall
much higher than the energy scale of FQH states in semiconductors.
\end{abstract}

\maketitle
\section{Introduction}

Landau symmetry breaking\cite{L3726,LanL58} has been the standard
theoretical concept in the classification of phases and transitions
between them. However, this theory turned out insufficient when the
fractional quantum Hall (FQH) state\cite{TSG8259,L8395} was discovered. These
states (FQH states and spin liquids) are not distinguished by their symmetries; instead they have new topological quantum numbers such as robust ground
state degeneracy\cite{Wtop,WNtop} and robust non-Abelian Berry's
phases\cite{WZ8411}. The topological
order\cite{Wrig,Wtoprev} associated with topological quantum numbers has been
proposed for the classification of these states. Recently, it was realized
that topological order can be interpreted as patterns of long range quantum
entanglement\cite{KP0604,LW0605,CGW1035}. This long range entanglement has
important applications for topological quantum computation: the robust ground state degeneracy can be used as quantum
memory\cite{DKL0252}; fractional defects from the entangled states which
carry fractional charges\cite{L8395} and fractional
statistics\cite{LM7701,W8257,ASW8422} (or non-Abelian
statistics\cite{MR9162,Wnab}) can perform fault tolerant quantum
computation\cite{K032,NSS0883}.

Although it has attractive concepts and applications, topological
order is only realized at very low temperatures in FQH
systems\cite{TSG8259,L8395}. In this paper we present a proposal to realize
highly entangled topological states at higher temperatures.  The ideal is to
combine geometric frustration, spin-orbit coupling and ferromagnetism in
quantum spin systems.  Both spin-orbit coupling and ferromagnetism can have
high energy scales and appear at room temperature.  Their combination breaks time-reversal symmetry which leads to rich and complicated
interference from quantum spin fluctuations.
In this paper we show that they can lead to
highly entangled topological states at high temperatures.

Quantum spins on the kagome lattice
are geometrically frustrated systems. They
appear in the following compounds: Herbertsmithite $\text{Zn Cu}_3\text{ (OH)}_6\text{ Cl}_2$,\cite{HMS0704,RML0915,HYO0902}
Kapellasite $\text{Cu}_3\text{Zn(OH)}_6\text{Cl}_2$,\cite{CRW0897}
$\text{Y}_{0.5}\text{Ca}_{0.5}\text{BaCo}_4\text{O}_7$,\cite{SEM1083}
$Mg_xCu_{4-x}(OH)_6Cl_2$,\cite{CMC1070}
$CaBaCo_4O_7$,\cite{CPH1044}
$Pr_3Ga_5SiO_{14}$,\cite{ZBM1002}
$Nd_3Ga_5SiO_{14}$,\cite{RSC0605}
$BaCu_3V_2O_8(OH)_2$,\cite{ZOO0935}
Cu(1,3-benzenedicarboxylate),\cite{MOK0987}
$KFe_3(OH)_6(SO_4)_2$,\cite{MHG0929}
$YBaCo_4O_7$,\cite{KMO1083,KMM1066}
$YBaCo_3AlO_7$,
$YBaCo_3FeO_7$,\cite{HHV0911,KMO1083}
$\gamma$-$Cu_2(OD)_3Cl$,\cite{WPR0956}
$Ni_{5}(TeO_{3})_{4}Br_{2}$,
$Ni_{5}(TeO_{3})_{4}Cl_{2}$,\cite{HMS0956}
$Cu_3V_2O_7(OH)_2/2H_2O$,\cite{YOT0904,YTY0907}
$Cs_2Cu_3CeF_{12}$,\cite{AYM0928}
$Cs_2Cu_3SnF_{12}$,
$Rb_2Cu_3SnF_{12}$,\cite{OMY0905}
$Cu_{2}(OD)_3Cl$,\cite{KJL0801}
$Cs_2Cu_3ZrF_{12}$,
$Cs_2Cu_3HfF_{12}$\cite{YOS0640} and
$Co_3V_2O_8$.\cite{PWB1010}
Motivated by these materials, in this paper we study the Heisenberg
model on the kagome lattice with additional spin-orbit interaction and Zeeman
coupling $\sum_i B_z S_i^z$.  Some related theoretical work can be found in Ref. 
\onlinecite{MCL1028,NS1028}.  In Ref. \onlinecite{MCL1028} a model with  spin-orbit interaction
but no Zeeman coupling is studied; some mean-field spin liquid states are
found. In Ref. \onlinecite{NS1028}  a model with Zeeman coupling but no spin-orbit
interaction is studied via numerical calculations.  Two magnetization steps are
found at $M/M_{\text{max}}=1/3$ (stronger) and $2/3$ (weaker) for a 36 spin cluster.

In this paper, we study the state with magnetization $\langle S_i^z\rangle=1/3$. In section \ref{sec:qs_ka_soc}, we write down the quantum spin model with spin-orbit coupling on the Kagome lattice. In section \ref{sec:top}, we map the spin model to the hardcore bosonic model in \ref{subsec:boson_model} , construct three trial wavefunctions for the polarized spin system $\langle S_i^z\rangle=1/3$ in \ref{subsec:wave} and then evaluate the energy expectation for these three states in \ref{subsec:num}. We find that the bosonic quantum Hall state has the lowest energy. Lastly, we discuss the materials realization in \ref{subsec:mat}. In the Appendix \ref{sec:spin_orbit}, we also discuss spin-orbit coupling in the transition metal oxide materials.

\section{Quantum spins on the kagome lattice with
spin-orbit coupling}\label{sec:qs_ka_soc}
\begin{figure}
  \includegraphics[width=5cm]{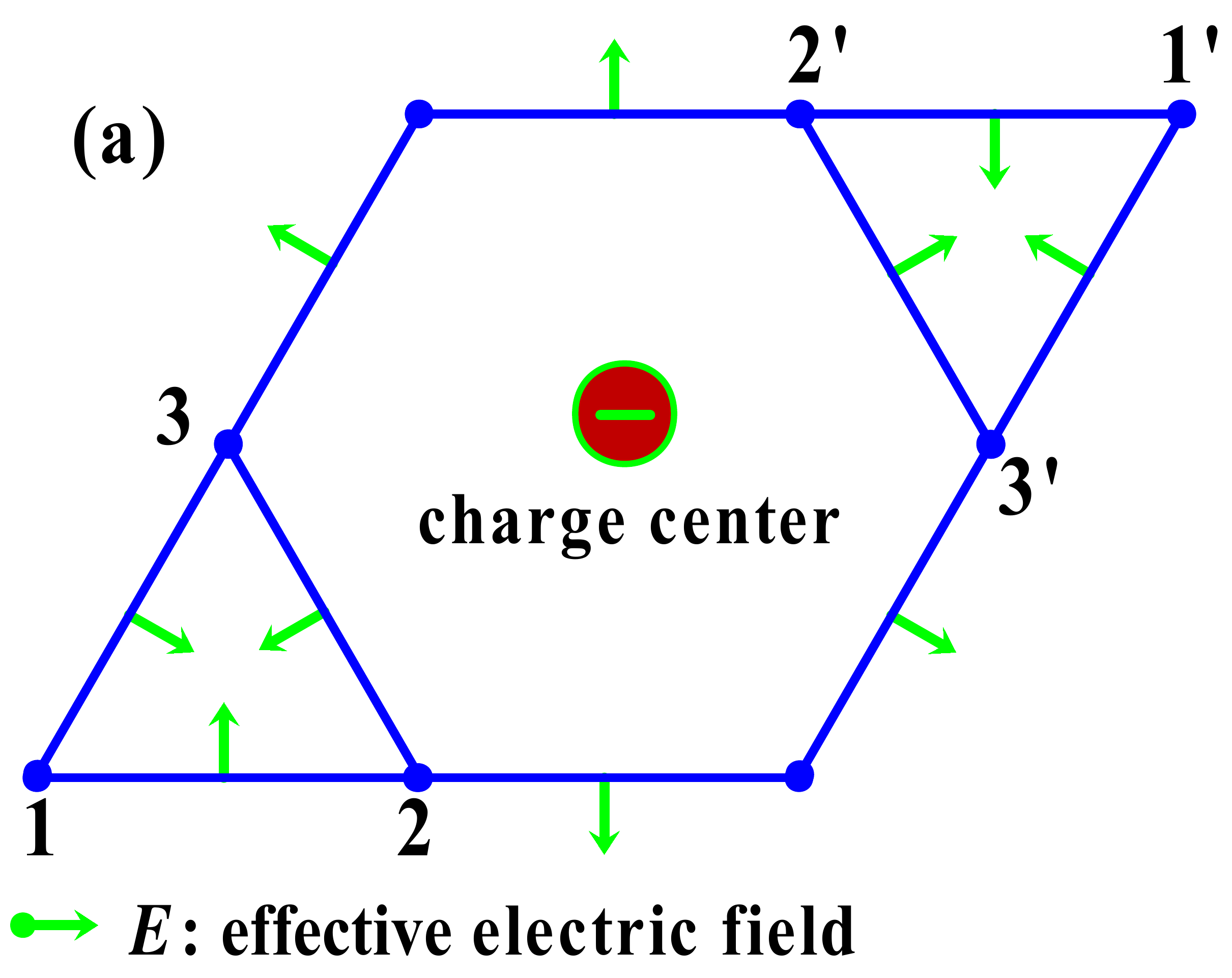}\\
  \includegraphics[width=\columnwidth]{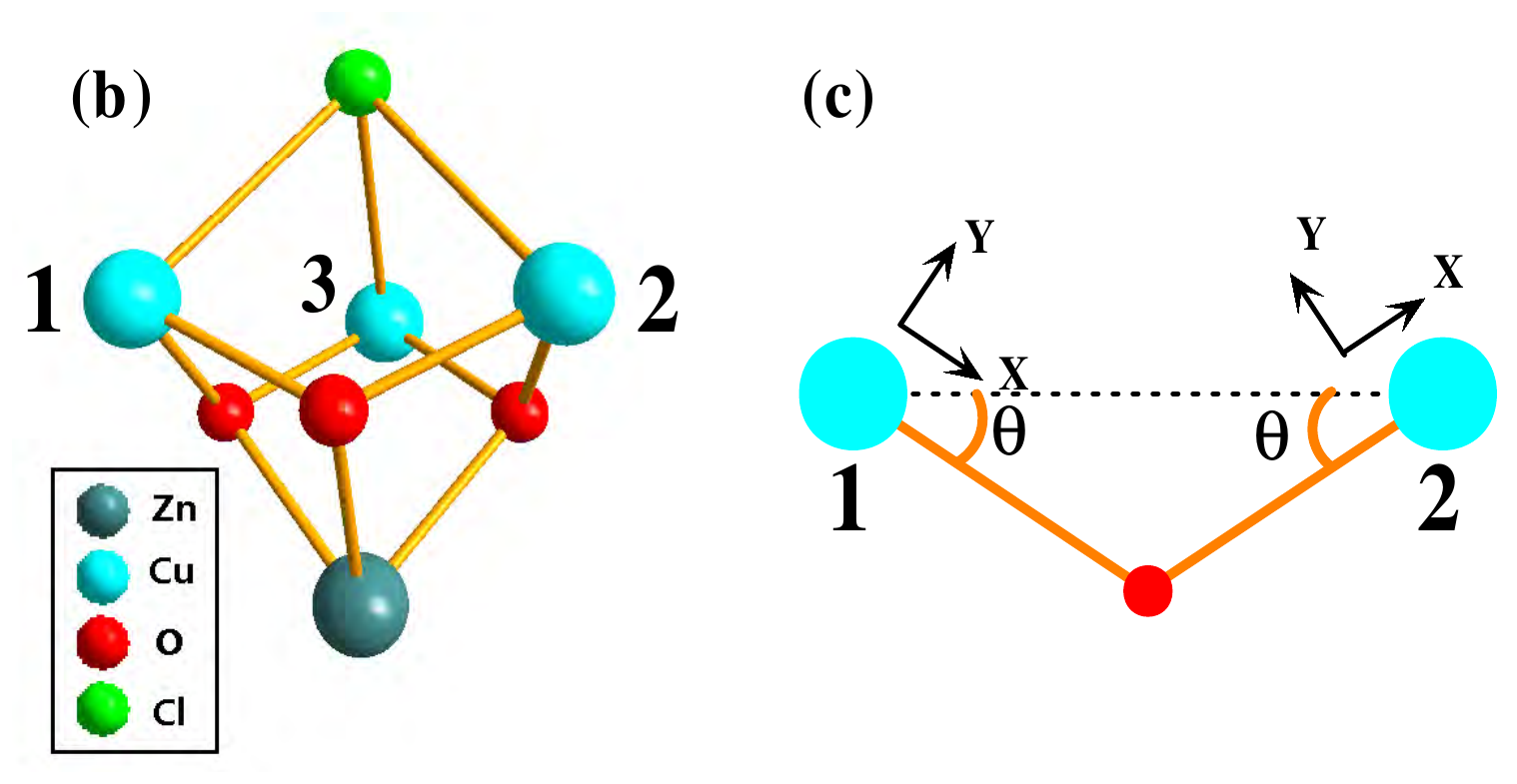}
  \caption{(color online) (a) The kagome lattice with three different  sites
$l=1,2,3$ within the unit cell. Inversion symmetry breaking via a
charge center in the hexagon leads to the effective electric field
$\mathbf{E}_{ij}$ on the bond $\mathbf{r}_{ij}$, represented by green arrows
which point from the middle of the bond to the center of every triangle on the
kagome lattice. (b) The triangle $\triangle_{123}$ in the kagome lattice for Herbertsmithite\cite{HMS0704,RML0915,HYO0902}. (c) The nearest neighbor bond $\mathbf{r}_{12}$: the electron hops from site
$\mathbf{r}_1$ to $\mathbf{r}_2$ mediated by the oxygen atom.}
  \label{fig:kagome}
\end{figure}

The kagome lattice has 3 sites (labelled $1$, $2$ and $3$) within every unit
cell with the primitive vectors $\mathbf{a}_1=2a\hat{\mathbf{x}}$ and
$\mathbf{a}_2=a(\hat{\mathbf{x}}+\sqrt{3}\hat{\mathbf{y}})$ ($a$ is the lattice
constant), see Fig. \ref{fig:kagome}(a). The unit cell contains one hexagon and
two triangles so it is geometrically frustrated.

As shown in Fig. \ref{fig:kagome} (b), the triangle $\triangle_{123}$ on the kagome plane in Herbertsmithite $\text{ZnCu}_3 \text{(OH)}_6 \text{Cl}_2$\cite{HMS0704,RML0915,HYO0902} contains three copper cations surrounded by distorted octahedrons sharing one chlorine corner while each pair shares an oxygen corner. Mediated by this oxygen, the Cu $3d^9$ electron hops from site $\mathbf{r}_1$ to $\mathbf{r}_2$, e.g. see Fig. \ref{fig:kagome}(c).

Inversion symmetry for Herbertsmithite breaks down explicitly, leading
to a non-uniform charge distribution in the kagome lattice. For convenience,
we model the  charge center as being in the hexagon, see Fig.\ref{fig:kagome}(a).  When
hopping from $\mathbf{r}_1$ to $\mathbf{r}_2$, the electron sees the electric
field $\mathbf{E}_{12}$ (labelled by the green arrow in Fig.
\ref{fig:kagome}(a)). The effective electric field couples to the electron
through the spin-orbit coupling vector
$\mathbf{D}_{12}=\alpha\mathbf{E}_{12}\times\mathbf{r}_{12}$ in the Rashba
manner, where $\mathbf{r}_{12}=\mathbf{r}_{1}-\mathbf{r}_{2}$\cite{Rashba1960}
\begin{eqnarray}
\label{eq:hoping}
  t_{12}=-t\sum_{\sigma\sigma'}\left(\exp(-i\vec{\sigma} \cdot\mathbf{D}_{12})_{\sigma\sigma'} c_{1\sigma}^\dag c_{2\sigma'}+\text{h.c.}\right)
\end{eqnarray}
Here $\vec{\sigma}=(\sigma_x,\sigma_y,\sigma_z)$ are the Pauli matrices. The
coefficient $\alpha$ should be chosen to make the spin-orbit
coupling vector $\mathbf{D}_{12}$ dimensionless. Note that
$\mathbf{D}_{12}=-\mathbf{D}_{12}$.

Including on-site interactions we obtain the Hubbard model with spin-orbit
coupling for $S=1/2$ electrons on the kagome lattice
\begin{eqnarray}
 H=-t\sum_{\sigma\sigma'}\left((e^{-i\vec{\sigma} \cdot\mathbf{D}_{ij}})_{\sigma\sigma'} c_{i\sigma}^\dag c_{j\sigma'}+\text{h.c.}\right)+U\sum_{i}n_{i\uparrow}n_{i\downarrow}
\end{eqnarray}
where $i$ and $j$ denote nearest neighbors.

For a specified bond $\mathbf{r}_{ij}$, we can make a gauge transformation \cite{PhysRevLett.69.836}
\begin{eqnarray}
\label{eq:gauge}
  c_{i\sigma}\rightarrow\tilde{c}_{i\sigma}&=& \sum_{\sigma'}(e^{i(D/2)\vec{\sigma} \cdot {\mathbf{n}}_{ij}})_{\sigma\sigma'}c_{i\sigma'}\nonumber\\
  c_{j\sigma}\rightarrow\tilde{c}_{j\sigma}&=& \sum_{\sigma'}(e^{-i(D/2)\vec{\sigma} \cdot {\mathbf{n}}_{ij}})_{\sigma\sigma'}c_{j\sigma'}
\end{eqnarray}
where $\mathbf{D}_{ij}=\mathbf{n}_{ij}D$. Then
\begin{eqnarray}
 H=-t\sum_{\sigma}(\tilde{c}_{i\sigma}^\dag \tilde{c}_{j\sigma}+\text{h.c.})+U\sum_{i}\tilde{n}_{i\uparrow}\tilde{n}_{i\downarrow}
\end{eqnarray}
Using standard second-order perturbation theory, we obtain the exchange term
\begin{eqnarray}
\label{eq:Jterm}
  J_{ij}=J\tilde{\mathbf{S}}_i\cdot\tilde{\mathbf{S}}_j
\end{eqnarray}
Here $J=4t^2/U$ is the exchange coupling for the rotated  spin operator
$\tilde{\mathbf{S}}_i=\sum_{\sigma\sigma'}\tilde{c}_{i\sigma}
\vec{\sigma}_{\sigma\sigma'} \tilde{c}_{j\sigma'}$.

On the kagome lattice, we cannot find a gauge
transformation as in Eq. (\ref{eq:gauge}) that would be compatible for each site. So we have to write the Hamiltonian in terms of the original spin operators. On every bond, the rotated spin operators are related
to the original ones as follows:
\begin{eqnarray}
  \tilde{S}_i&=&(1-\cos(D))(\hat{\mathbf{n}}_{ij}\cdot \mathbf{S}_i)\hat{\mathbf{n}}_{ij}+\cos(D)\mathbf{S}_i\nonumber\\ &&-\sin(D)\mathbf{S}_i\times\hat{\mathbf{n}}_{ij}\\
  \tilde{S}_j&=&(1-\cos(D))(\hat{\mathbf{n}}_{ij}\cdot \mathbf{S}_j)\hat{\mathbf{n}}_{ij}+\cos(D)\mathbf{S}_j\nonumber\\ &&+\sin(D)\mathbf{S}_j\times\hat{\mathbf{n}}_{ij}
\end{eqnarray}
Thus we obtain the quantum spin model on the kagome lattice including
spin-orbit coupling\cite{PhysRevLett.69.836}
\begin{eqnarray}
\label{spinmodel}
  H=&&J\sum_{\langle ij\rangle}\left(\cos(2D)\mathbf{S}_i\cdot \mathbf{S}_j+\sin(2D)(\mathbf{S}_i\times \mathbf{S}_j)\cdot\hat{\mathbf{n}}_{ij}\right.\nonumber\\
  &&\left.+2\sin^2(D)(\mathbf{S}_i\cdot\hat{\mathbf{n}}_{ij})(S_j\cdot\hat{\mathbf{n}}_{ij})\right).
\end{eqnarray}

\section{Polarized spin state  with topological order}\label{sec:top}

\subsection{Hardcore bosonic model}\label{subsec:boson_model}

\begin{figure}
 \includegraphics[width=5cm]{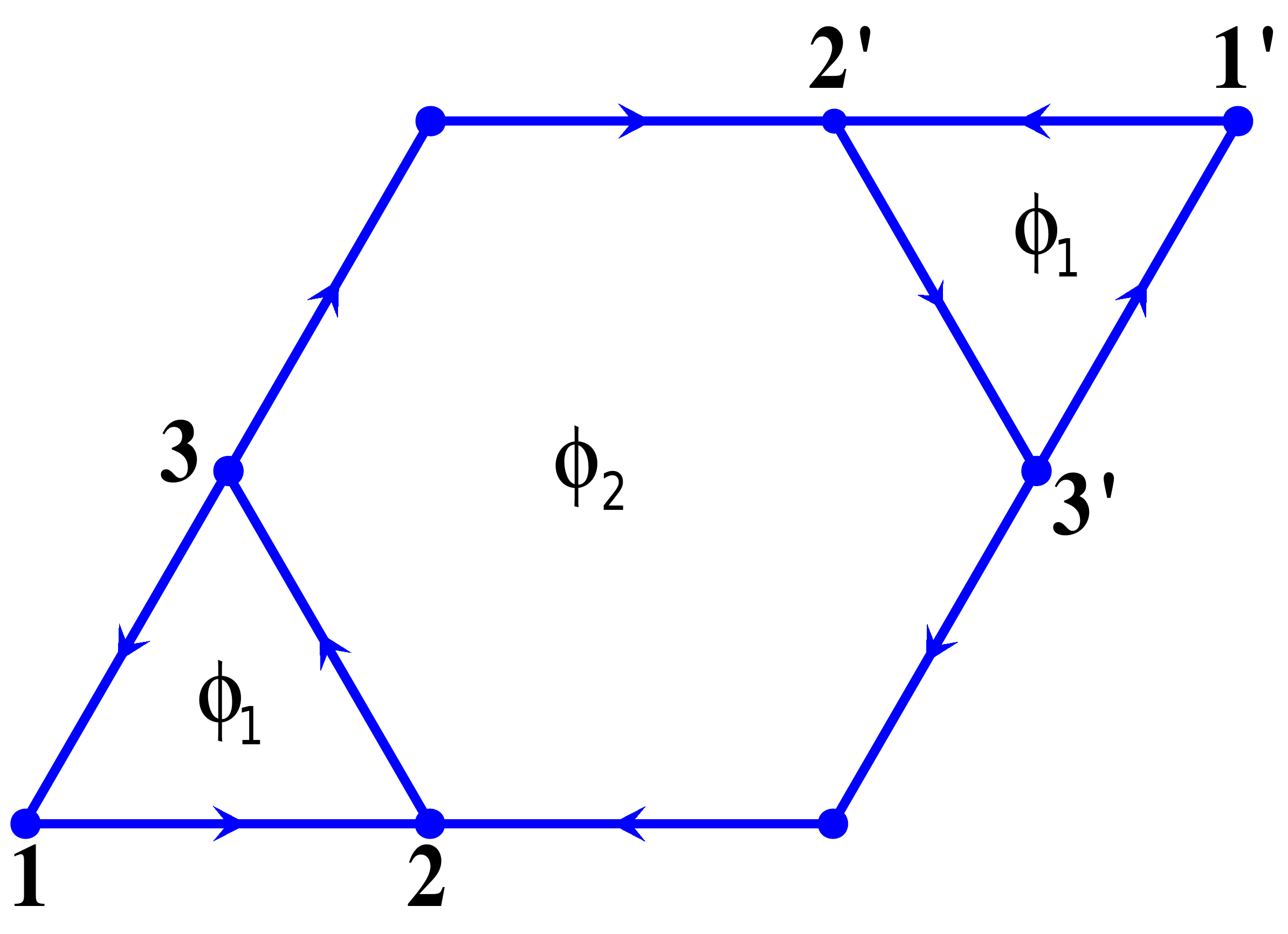}
  \caption{(color online) Flux distribution for the hardcore bosons.}\label{fig:flux}
\end{figure}

For simplicity we choose the spin-orbit coupling vectors
$\mathbf{D}_{ij}$ perpendicular to the kagome plane:
$\mathbf{D}_{12}=\mathbf{D}_{23}=\mathbf{D}_{31}=\mathbf{D}_{1'2'}=\mathbf{D}_{2'3'}=\mathbf{D}_{3'1'}
=D\hat{\mathbf{z}}$ (only in-plane effective electric fields $\mathbf{E}_{ij}$ are considered). We use  the Holstein-Primakoff transformation:
\begin{eqnarray}
  S_i^{+}=b_i^\dag\quad, S_i^{-}=b_i\quad, S_i^{z}={1\over2}-b_i^\dag b_i
\end{eqnarray}
where $b_i$ is the hardcore bosonic operator $  [b_i,b_j^\dag]=\delta_{ij}$,
$n_i=b_i^\dag b_i\le1$. This maps the spin model (\ref{spinmodel}) onto a
hardcore bosonic model\cite{PhysRevLett.59.2095,  PhysRevLett.70.2641}
\begin{eqnarray}
\label{eq:bosonic}
  H&=&{J\over2}\sum_{\langle ij\rangle}\left( \exp[(\hat{\mathbf{n}}_{ij}\cdot\hat{\mathbf{z}})i2D]b_i^\dag b_j+\text{h.c.}\right)\nonumber\\
  &&+J\sum_{\langle ij\rangle}n_in_j
\end{eqnarray}
which describes interacting hardcore systems with hopping under
effective fluxes as shown in Fig. \ref{fig:flux}: within the triangles
$\triangle_{123}$ and $\nabla_{1'2'3'}$, there are fluxes $\phi_1=6D$; in the
hexagon, there is flux $\phi_2=-2\phi_1$.    When
$\phi_1\neq 0,\pi\mod 2 \pi$, these effective fluxes break time-reversal symmetry for this model.

\begin{figure}
 \includegraphics[width=\columnwidth]{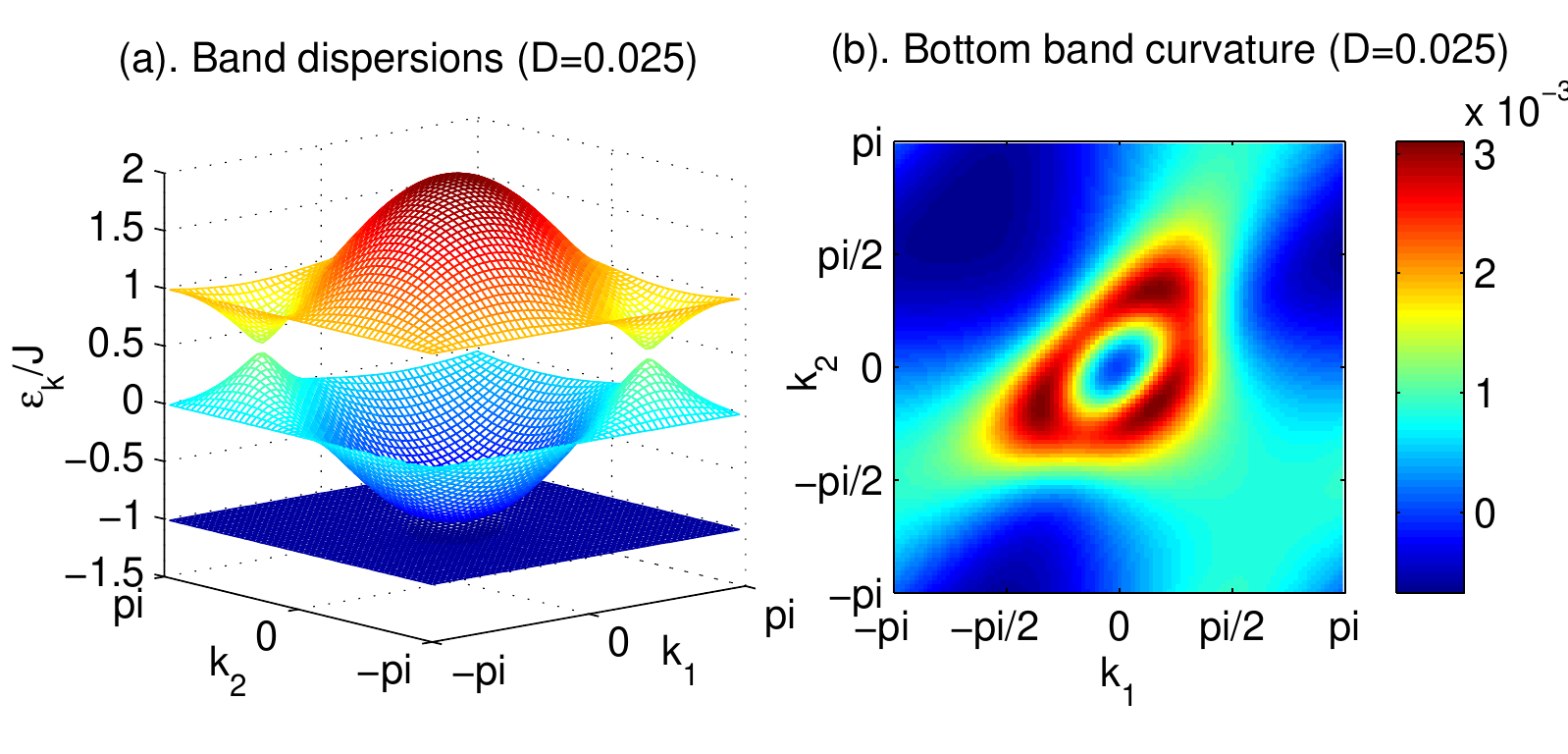}\\ \includegraphics[width=\columnwidth]{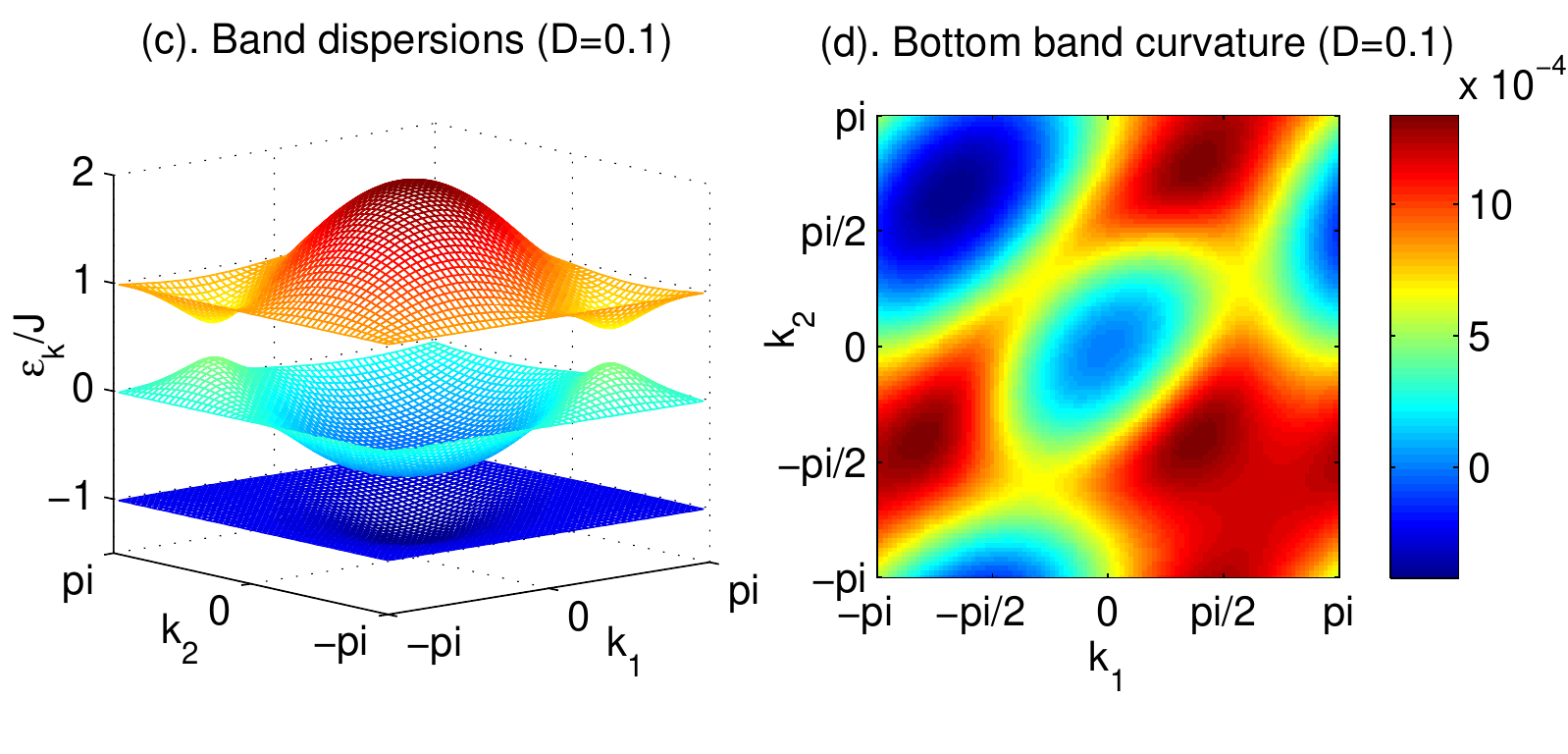}\\ \includegraphics[width=\columnwidth]{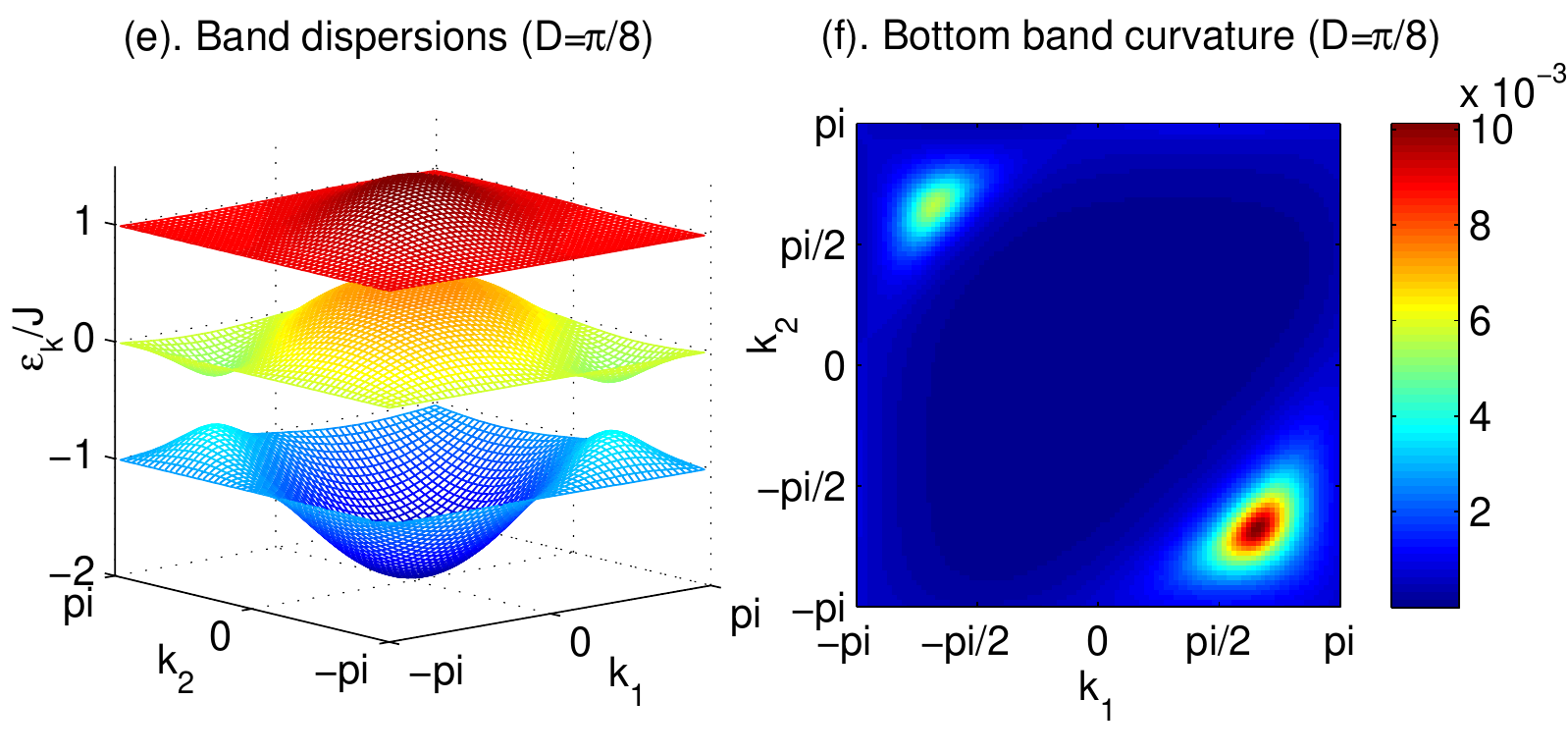}\\
  \caption{(color online) (a), (c)  and (e): Band dispersions of the hardcore boson in the presence of the flux as shown in Fig. \ref{fig:flux} for $D=0.025$, $D=0.1$ and $D=\pi/8$, respectively; (b), (d) and (f) are the corresponding bottom band curvatures for (a), (c) and (e). }\label{fig:curv}
\end{figure}

Now let us consider just one boson described by only the hopping term in the above
Hamiltonian.  The hopping  Hamiltonian has three bands.  We calculate the Berry
curvatures over the Brillouin zone for the lowest band in the presence of the flux for different
spin-orbit couplings: $D=0.025$, $0.1$ and $\pi/8$ . The
Berry curvature is defined as follows
\begin{eqnarray}
  F_n(\mathbf{k})=\epsilon_{ij}\partial_{k_i} A_j^{(n)}(\mathbf{k}),~
  A_i^{(n)}(\mathbf{k})=i\langle u_{n\mathbf{k}}|\partial_{k_i}|u_{n\mathbf{k}}\rangle
\end{eqnarray}
where $u_{n\mathbf{k}}$ is the Bloch wave packet in the $n$-th band of the hopping Hamiltonian
\begin{eqnarray}
  H_t={J\over2}\sum_{\langle ij\rangle}\left( \exp[(\hat{\mathbf{n}}_{ij}\cdot\hat{\mathbf{z}})i2D]b_i^\dag b_j+\text{h.c.}\right)\nonumber\\
  H_t |u_{n\mathbf{k}}\rangle=\epsilon_{n\mathbf{k}}|u_{n\mathbf{k}}\rangle
\end{eqnarray}
In this paper, instead of using $\hat{k}_x$ and $\hat{k}_y$ as the axes in $k$-space, we use $\hat{k}_1=k_x$ and
$\hat{k}_2=(\hat{k}_x+\sqrt{3}\hat{k}_y)/2$ for convenience. The dispersion
$\epsilon_{n\mathbf{k}}$ has three bands (labelled $n=b$ for the bottom band,
$n=m$ for the middle band and $n=t$ for the top band) as shown in Fig.
\ref{fig:curv} (a), (c) and (e).  In all cases ($D=0.025$, $D=0.1$ and $D=\pi/8$) the three bands  have  nonzero Chern numbers $C_{\text{b}}=1$, $C_{\text{m}}=0$ and $C_{\text{t}}=-1$, where the Chern number $C\equiv{1\over2\pi}\int_{\text{BZ}}d^2kF_n(\mathbf{k})$.

We plot the Berry curvature of the bottom band for $D=0.025$, $0.1$ and $\pi/8$
in Fig. \ref{fig:curv} (b), (d) and (f).  We see that when $D=0.1$, the lowest
band is separated from the other bands by an energy gap and the lowest band is
quite flat.  Since the lowest band has a non-zero Chern number $C_b=1$, it
simulates the first Landau level in free space.  By analogy to the quantum
Hall effect in high magnetic field, the hardcore bosons are likely to form a
$\nu=1/2$ bosonic quantum Hall state when there is half a boson per unit
cell. The boson filling number is $f=1/6$ per site which corresponds to the
spin polarization
\begin{eqnarray}
  \langle S_i^z\rangle=1/2-f=1/3
\end{eqnarray}
In other words, the polarized spin state $\langle S_i^z\rangle=1/3$ is likely
to be a chiral spin liquid\cite{WWZcsp} --- a topologically ordered state.

\subsection{Fermionic constructions for the bosonic wavefunctions}\label{subsec:wave}

To study the topologically ordered chiral spin state, we will employ the
fermionic approach to construct trial bosonic wavefunctions . The many-body
bosonic wavefunction can be represented as follows
\begin{eqnarray}
  |\Phi\rangle=\sum_{\{x_1,\cdots,x_{N_b}\}}\Phi(x_1,\cdots,x_{N_b})|\{x_i\}\rangle
\end{eqnarray}
Here the sum is over all possible boson configurations
$|\{x_i\}\rangle=|\{x_1,\cdots,x_{N_b}\}\rangle=a_{x_1}^\dag \cdots
a_{x_{N_b}}^\dag |0\rangle$ and  $\Phi(x_1,\cdots,x_{N_b})$ is the symmetric
wavefunction. In this paper, we are only concerned with translationally
invariant ground states.

The many-body bosonic wave function for a product state (PS)
has the form
\begin{eqnarray}
  \label{eq:ps}
  \Phi(x_1,\cdots,x_{N_b})=\prod_{i=1}^N\phi_l(x_i)
\end{eqnarray}
where $\phi_l(x_i+\mathbf{a}_j)=\phi_l(x_i)$ to maintain translational
invariance. $l=1,2,3$ denotes sites within the unit cell and
the vector $\mathbf{a}_j$, $j=1,2$, is a Bravais vector for the kagome lattice.
So the many-body wave function is labelled by three complex parameters $\phi_1$,
$\phi_2$, and $\phi_3$ corresponding to the spin orientation on the three
sites in each unit cell.  The mean field ground state of this type is obtained
by minimizing the average energy by varying $\phi_1$, $\phi_2$, and $\phi_3$.
This type of spin ordered states without topological
order is the main competing state for the ground state of our model.

To construct the bosonic ground state with topological order, we split the the
hardcore boson into two species of fermions
\begin{eqnarray}
\label{eq:frac}
  b_i=\alpha_i\beta_i
\end{eqnarray}
where $\alpha_i$ and $\beta_i$ are fermion operators which satisfy the
constraint on every site: $n_{i\alpha}=n_{i\beta}=n_{ib}$. The configuration becomes
\begin{eqnarray}
  |\{x_i\}\rangle=\alpha_{x_1}^\dag\beta_{x_1}^\dag \cdots \alpha_{x_{N_b}}^\dag\beta_{x_{N_b}}^\dag  |0\rangle
\end{eqnarray}
and the symmetric wave function factorizes as follows
\begin{eqnarray}
  \Phi(\{x_i\})=\Psi_\alpha(\{x_i\})  \Psi_\beta(\{x_i\})
  \label{eq:factorize}
\end{eqnarray}
where $\Psi_\alpha(\{x_i\})$ and $\Psi_\beta(\{x_i\})$ are the antisymmetric
fermionic wavefunctions for $\alpha_i$ and $\beta_i$. Using this
fractionalization we construct two ansatz wavefunctions: the bosonic
quantum Hall state (QHS) and the spin Hall state (SHS).

The fermionic wavefunctions $\Psi_\alpha(\{x_i\})$ and $\Psi_\beta(\{x_i\})$ can be constructed from the mean field tight binding Hamiltonian
\begin{eqnarray}
\label{eq:mean_field}
  H_\alpha&=&-t_\text{eff}\sum_{\langle ij\rangle}\left(\alpha_i^\dag\alpha_j\exp(iA_{ij}^\alpha) +\text{h.c.}\right)\nonumber\\
  H_\beta&=&-t_\text{eff}\sum_{\langle ij\rangle}\left(\beta_i^\dag\beta_j\exp(iA_{ij}^\beta) +\text{h.c.}\right)
\end{eqnarray}
The filling factors for the fermions $\alpha_i$,$\beta_i$ are $f=1/6$ per site,
namely half per unit cell. For the QHS and SHS, we need the filling factor
corresponding to one particle per unit cell which can be realized by inserting
half a flux quantum ($\phi=\pi$) in the original unit cell to double the unit
cell:
\begin{eqnarray}
  \phi_2^\omega+2\phi_1^\omega=\pi,\quad \omega=\alpha,\beta
\end{eqnarray}
where $\phi_1^\omega$ and $\phi_2^\omega$ are fluxes in the kagome unit cell,
see Fig. \ref{fig:flux}. In the presence of these fluxes, we specify a gauge
for $A_{ij}^\omega$ in this tight-binding model (\ref{eq:mean_field}) to obtain
single particle wavefunctions in the bottom band:
$\psi_\omega(\mathbf{k}_i,x_j)$ ($\omega=\alpha,\beta$), where $\mathbf{k}$ is
the Bloch momentum vector for the doubled unit cell. Thus the fermionic
wavefunctions are the determinants of these single particle wave functions:
\begin{eqnarray}
  \Psi_\omega(\{x_i\})&=&\det[\psi_\omega(\mathbf{k}_i,x_j)]
\end{eqnarray}
where $\omega=\alpha,\beta$ and $i,j=1,2,\cdots,N_b$. For the QHS state, we set
$\phi_1^\alpha=\phi_1^\beta$ and $\phi_2^\alpha=\phi_2^\beta$; for the SHS
state, we set $\phi_1^\alpha=-\phi_1^\beta$ and $\phi_2^\alpha=-\phi_2^\beta$.
For the QHS, each fermion has the same Chern number $C_\omega=1$; for
the SHS, $C_\alpha=1$ and $C_\beta=-1$.

We now consider the effective theory for the QHS and SHS.
%For the PS, the low energy dynamics can be described by the phase
%fluctuations in the 2+1D $XY$ model
%\begin{eqnarray}
%  \mathcal{L}_{\text{PS}}={1\over g_1}\sum_\mu(\partial_\mu\varphi)^2
%\end{eqnarray}
%Introduce the gauge field $a_\mu$ to describe the current operator
%\begin{eqnarray}
%  j_\mu=\partial_\mu\varphi={1\over2\pi}\epsilon_{\mu\nu\lambda}\partial_\nu a_\lambda
%\end{eqnarray}
%we dual the the 2+1D $XY$ model to the 2+1D $U(1)$ theory
%\begin{eqnarray}
%  \mathcal{L}_{\text{PS}}= {1\over 4\pi^2g_1}(\epsilon_{\mu\nu\lambda}\partial_\nu a_\lambda)^2
%\end{eqnarray}
%
For the QHS and SHS, fermionic excitations are gapped out. There is a gauge
freedom for the fractionalization in Eq. (\ref{eq:frac}): the gauge
transformation $\alpha_i\rightarrow\alpha_i e^{i\theta_i}$,
$\beta_i\rightarrow\beta_i e^{-i\theta_i}$ does not change the bosonic operator
$b_i$.
%The gauge fluctuation associated is now
%$\tilde{A}_{ij}=\theta_j-\theta_i$.
The Hamiltonian with the gauge fluctuations is given by
\begin{eqnarray}
  H_\alpha&=&-t_\text{eff}\sum_{\langle ij\rangle}\left(\alpha_i^\dag\alpha_j\exp(i\tilde{A}_{ij}+ia_{ij}) +\text{h.c.}\right)\nonumber\\
  H_\beta&=&-t_\text{eff}\sum_{\langle ij\rangle}\left(\beta_i^\dag\beta_j\exp(-i\tilde{A}_{ij}-ia_{ij}) +\text{h.c.}\right)
\end{eqnarray}
%Introduce $U(1)$ gauge field $a_\mu^\omega$ to describe the fermionic currents
%\begin{eqnarray}
%  j_\mu^\omega={1\over2\pi}\epsilon_{\mu\nu\lambda}\partial_\nu a_\lambda^\omega
%\end{eqnarray}
The non-zero Chern number for each fermion species implies that the low
energy effective action for the gauge fields is given by
\begin{eqnarray}
  \mathcal{L}&=&{i\over4\pi}\sum_{\omega}C_\omega\epsilon_{\mu\nu\lambda} a_\mu\partial_\nu a_\lambda
+ ...
%  &&+{i\over2\pi}\epsilon_{\mu\nu\lambda} (C_\alpha a_\mu+C_\beta a_\mu)\partial_\nu \tilde{A}_\lambda
\end{eqnarray}
%Since $n_{i\alpha}=n_{i\beta}=n_{ia}$, we have the constraint
%\begin{eqnarray}
%  a_\mu=a_\mu^\alpha=a_\mu^\beta
%\end{eqnarray}
where $...$ represents higher order terms.
For the QHS, $C_\alpha=C_\beta=1$ so we obtain the low-energy effective theory
\begin{eqnarray}
  \mathcal{L}_\text{QHS}={i\over2\pi}\epsilon_{\mu\nu\lambda} a_\mu\partial_\nu a_\lambda
+{1\over 4\pi^2g}(\epsilon_{\mu\nu\lambda}\partial_\nu a_\lambda)^2
\end{eqnarray}
This describes the $\nu=1/2$ FQH state
for bosons corresponding to the chiral spin state first introduced in
Ref. \onlinecite{WWZcsp}.  We note that although the $\alpha$ and $\beta$
fermions have the same Chern number, the sign of the coupling of each fermion to the $U(1)$ gauge field $a_\mu$ is opposite.
Thus a $2\pi$ flux of $a_\mu$ creates an $\alpha$ fermion
and annihilates a $\beta$ fermion.

For the SHS, $C_\alpha=1$, $C_\beta=-1$ and
we obtain the low-energy effective theory
\begin{eqnarray}
  \mathcal{L}_\text{SHS}={1\over 4\pi^2g}(\epsilon_{\mu\nu\lambda}\partial_\nu a_\lambda)^2
\end{eqnarray}
Here the $\alpha$ and $\beta$ fermions in the SHS have opposite Chern number and the
sign of the coupling of the two fermions to the $U(1)$ gauge field $a_\mu$ is
also opposite.  Thus a $2\pi$ flux of $a_\mu$ creates an $\alpha$ fermion and a
$\beta$ fermion which corresponds to a $b$ boson, i.e. it describes a spin flip.  So the magnetic field of the  $U(1)$ gauge field
$a_\mu$ corresponds to the spin $S^z$ density.  Since spin $S^z$ is conserved,
the $U(1)$ instanton is forbidden. Thus the  2+1D $U(1)$ gauge theory above is
not confined and the  $U(1)$ gauge field $a_\mu$ remains gapless. As this
gapless $U(1)$ gauge field corresponds to the spin $S^z$ density, the gapless
spin density fluctuations imply that $e^{i\theta S^z}$ spin rotation is
spontaneously broken.  Thus the SHS is a spin XY ordered state.

\subsection{Numerical results}\label{subsec:num}

\begin{figure}
  \includegraphics[width=6cm]{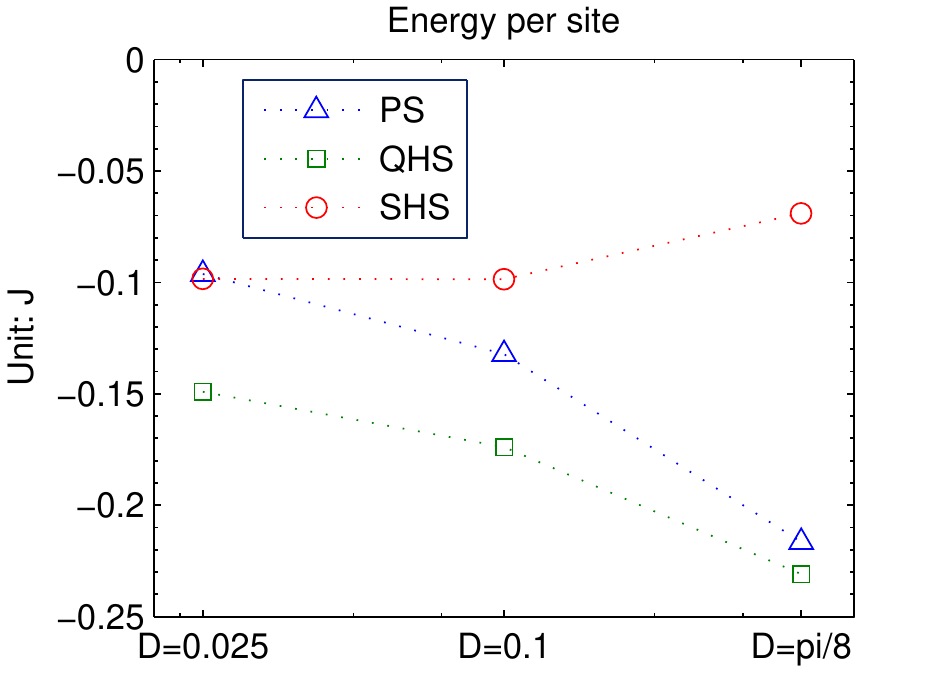}\\
  \caption{(color online) Energy per site for $D=0.025$, $D=0.1$ and $D=\pi/8$ respectively.}\label{fig:energy}
\end{figure}
For the three different states (PS, QHS and SHS), we can evaluate the expected energies for the bosonic model (\ref{eq:bosonic}):
\begin{eqnarray}
\label{eq:energy}
  E(\Phi)={\langle\Phi|H|\Phi\rangle\over\langle\Phi|\Phi\rangle}
  =\sum_{\{x_i\}}e_L(\{x_i\}) {|\langle\{x_i\}|\Phi\rangle|^2\over\langle\Phi|\Phi\rangle}
\end{eqnarray}
where we define the local energy
$e_L(\{x_i\})={\langle\Phi|H|\{x_i\}\rangle\over\langle\Phi|\{x_i\}\rangle}$.
We evaluate the energy in (\ref{eq:energy}) by appropriately averaging the
local energy $e_L(\{x_i\})$ over a set of configurations $|\{x_i\}\rangle$
distributed according to the square of the wave function
$|\langle\{x_i\}|\Phi\rangle|^2$, generated with a standard variational Monte
Carlo method.

Then we use a minimization function to optimize the expectation values  on an
$8\times8$ lattice.  In Fig. \ref{fig:energy}, we plot the energy per site of
the three states for $D=0.025$, $D=0.1$ and $D=\pi/8$. For $D=0.025$, the
PS and SHS have energies close to each other. The QHS has a better energy. As
the spin-orbit coupling is increased, the SHS becomes worse in energy. Both the PS
and QHS gain in energy and the PS gains much more. When $D=\pi/8$, the PS gives results close to the QHS. With small spin-orbit coupling ($D=0.025$ and $D=0.1$),
the bottom band of the hopping Hamiltonian in (\ref{eq:bosonic}) is flat and
has a smooth curvature over the Brillouin zone. The classic PS cannot gain much
energy through condensation of the lowest states. When $D$ increases, the bottom band becomes more convex and the PS will gain a lot of energy through condensation.

Our numerical results indicate that the topologically ordered QHS (the chiral
spin state) is a serious candidate for a kagome spin system with spin-orbit
coupling and spin polarization. This may be a realistic route for the discovery of new
topologically ordered states in quantum spin systems.

\subsection{Practical realization}\label{subsec:mat}

\begin{figure}
  \includegraphics[width=6cm]{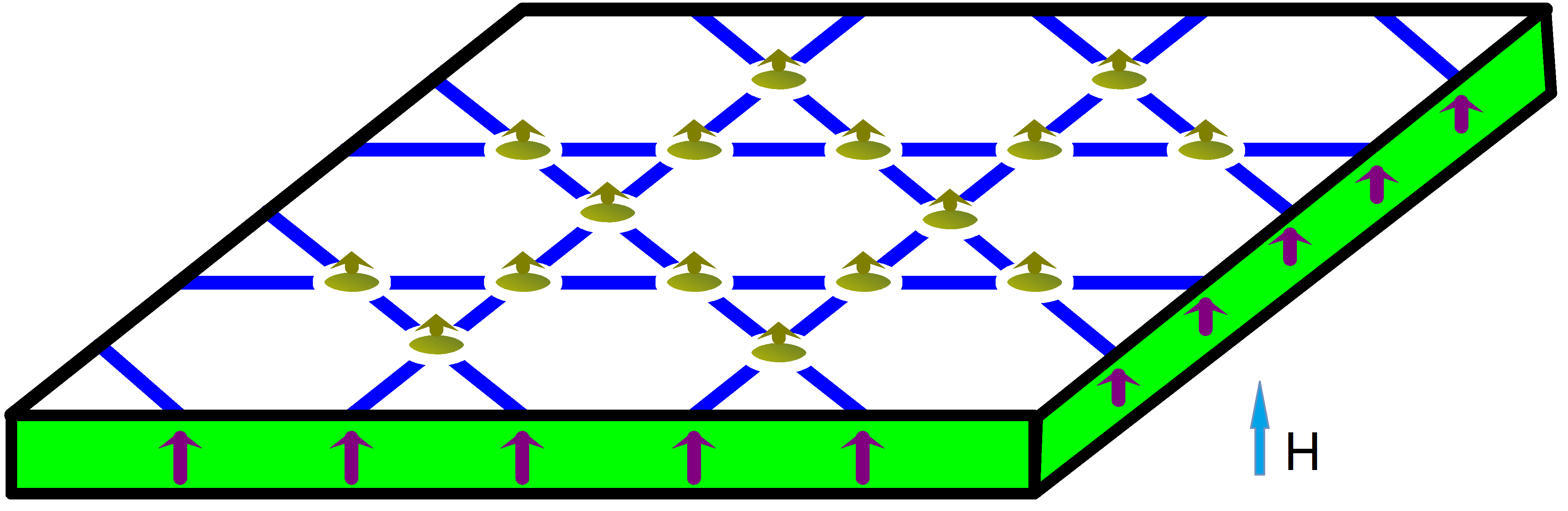}\\
  \caption{(color online) A scheme to tune the filling number of hardcore bosons $b_i$: the kagome lattice couples to a ferromagnetic substrate by the exchange interaction $H_{\text{int}}=J_{\text{ex}}\sum_{i}\mathbf{S}_i^m\cdot\mathbf{S}_{i}$; we can tune the substrate magnetization $\langle\mathbf{S}_i^m\rangle$ by  an applied magnetic field.  }\label{fig:kagome_mag}
\end{figure}

Here we explore the possibility of obtaining a polarized state ($S_i^z\neq 0$) experimentally. This can be achieved by applying a magnetic field, which adds a
term $H_h=-B\sum_in_i$ to the Hamiltonian (\ref{eq:bosonic}). However, the
exchange energy $J\sim 100 \text{meV}$ is usually very large, so experimentally accessible magnetic fields cannot polarize the spin to $S^z=1/3$. Hence we should find other ways of obtaining a large effective magnetic field.

One way is to place the kagome lattice on a ferromagnetic substrate, see Fig.
\ref{fig:kagome_mag}.  The exchange interaction is
$H_{\text{int}}=J_{\text{ex}}\sum_{i}\mathbf{S}_i^m\cdot\mathbf{S}_{i}$, here
$J_{\text{ex}}$ is the exchange coupling between spins $\mathbf{S}_i^m$ on the
substrate and spins $\mathbf{S}_{i}$ on the kagome plane. The exchange coupling
$J_{\text{eff}}$ can be very large when the kagome plane structure matches that
of the substrate perfectly.
%For the ferromagnetic substrate, we can use a small magnetic field $H$ to tune
%its magnetization $\langle S_i^{(m)z}\rangle$ and then we have the large
%effective magnetic field $B_\text{eff}=J_{\text{ex}}\langle S_i^{(m)z}\rangle$
%which is comparable with the exchange coupling on the kagome lattice.
Such a large effective magnetic field can polarize the spin on the kagome
lattice.

A third way is to insert ferromagnetic atoms in the kagome system. If these
ferromagnetic atoms form a ferromagnetic state, the exchange interaction can
also induce  spin polarization on the kagome lattice.

\section{Summary}

In this paper, we study quantum spin systems on the kagome lattice with
spin-orbit coupling and and spin polarization.  We argue that such a system can
be in a topologically ordered chiral spin state, a FQH state for bosonic spin
degrees of freedom.  The energy scale of the bosonic FQH state is of the same
order as the spin-orbit coupling and ferromagnetism --- overall much higher
than the energy scale of FQH states in semiconductors.  This result suggests
exploration of topologically ordered states in quantum spin systems with a
proper combination of geometric frustration, spin-orbital coupling and
ferromagnetism.

This research is supported by  NSF Grant No.  DMR-1005541 and NSFC 11074140.

\appendix

\section{Discussion of spin-orbit coupling}\label{sec:spin_orbit}

The Rashba spin-orbit coupling is weak, around $|D|=0.025$ for
Herbertsimthsite $\text{Zn Cu}_3 \text{(OH)}_6 \text{Cl}_2$. This small value prompts us to find other mechanisms to increase the strength of the
spin-orbit coupling.

In the literature, the spin-orbit coupling is also discussed on the atomic level and with a large strength of coupling, e.g. around 0.2 eV and 0.4 eV for the $4d$ and $5d$ electrons, respectively.  We hope to relate the atomic spin-orbit coupling to the form represented in Eq. \ref{eq:hoping}.  This has been achieved for the $5d^5$ electron in $\text{Sr}_2\text{IrO}_4$\cite{Wang2010} and we  discuss the general case below.

A $S=1/2$ electron can be found in $d^1$, $d^5$ and $d^9$ orbtitals in transition metal cations, e.g. in $\text{Mo}^{5+}$, $\text{Ir}^{4+}$ and $\text{Cu}^{2+}$ respectively.  Due to the crystal field , the fivefold degenerate $d$ state is split into a doublet $e_g$ and a triplet  $t_{2g}$.  $d^1$ and $d^5$ with a ligand octahedron and $d^9$ with a ligand tetrahedron belong to $t_{2g}$ . The triplet $t_{2g}$ has strong spin-orbit coupling
\begin{eqnarray}
\label{eq:soc}
 H_{i}=\lambda \mathbf{l}_{i}\cdot\mathbf{s}_{i}.
\end{eqnarray}
Here $\mathbf{s}_{i}$ is the spin operator and $l=1$ is the effective angular momentum with  $|l_i^{z}=0\rangle\equiv|XY\rangle_i$  and $|l_i^{z}=\pm1\rangle\equiv-{1\over\sqrt{2}}(i|XZ\rangle_i\pm|YZ\rangle_i)$ ($X$, $Y$ and $Z$ are local axes supporting by the local octahedron or tetrahedron , e.g. see Fig. \ref{fig:kagome}(b)).  The strong spin-orbit coupling (\ref{eq:soc}) splits $t_{2g}$ into two groups with effective angular momentum $J_\text{eff}^z=1/2$ and $J_\text{eff}^z=3/2$, respectively.  The $J_\text{eff}^z=1/2$ singlet contains a Kramers doublet: %$|\tilde{\sigma}\rangle_i={\sqrt{2}\over3}|l_z=-\sigma,\sigma\rangle_i+{1\over3}|0,-\sigma\rangle_i$, where $\sigma=\uparrow,\downarrow$ for the spin components corresponding  to $\sigma=\pm1$ in the $\hat{z}$ direction orbit components $l_z$, e.g.
$|\tilde{\uparrow}\rangle_i={\sqrt{2}\over3}|-1,\uparrow\rangle_i+{1\over3}|0,\downarrow\rangle_i$ and $|\tilde{\downarrow}\rangle_i={\sqrt{2}\over3}|+1,\downarrow\rangle_i+{1\over3}|0,\uparrow\rangle_i$. Here we are concerned with the Kramers doublet labelled by the pseudospin  $\tilde{\mathbf{s}}_i=\tilde{\mathbf{\sigma}}_i/2$.

There is a $p$ bond between the cation and mediating oxygen atom. When the two cations and oxygen lie on a straight line along the bond, the values of the overlap for $\langle l_z=0| p\rangle_\text{O}$ and $\langle l_z=\pm1|p\rangle_\text{O}$ are the same. In Herbertsmithite, they are not along a straight line and form a triangle, see Fig. \ref{fig:kagome}(c). As a result,   different orbits have different overlaps with the $p$ orbital  in the oxygen $\langle\tilde{\sigma}|p\rangle_\text{O}$ when the electron hops on the bond $\mathbf{r}_{12}$ from $\mathbf{r}_1$ to $\mathbf{r}_2$:
\begin{eqnarray}
|\langle l_z=\pm1|p\rangle_\text{O}|>|\langle l_z=0|p\rangle_\text{O}|
\end{eqnarray}
 For the sake of simplicity, we neglect the overlap $ |\langle l_z=0|p\rangle_\text{O}|$ here and are only concerned with $|\langle l_z=\pm1|p\rangle_\text{O}|$. Then we can write the spin-orbit in terms of the pseudospin in this manner
\begin{eqnarray}
\label{eq:soc1}
 H_{i}\sim\lambda \mathbf{l}(\mathbf{r}_i)\cdot\tilde{\mathbf{s}}(\mathbf{r}_i)
\end{eqnarray}
Here the orbital angular momentum $\mathbf{l}(\mathbf{r}_i)$ can be regarded as the effective magnetic field on the pseudospin $\tilde{\mathbf{s}}(\mathbf{r}_i)$. It is very interesting that the orientation of the \textit{magnetic field} $\mathbf{l}(\mathbf{r}_i)$ varies from each site $\mathbf{r}_i$. When the hopping encloses a loop, e.g. $\triangle_{123}$ in Fig. \ref{fig:kagome} (b), an electron obtains a non-zero Berry phase $\phi_\text{Berry}$ related to the spin-orbit coupling vector $\mathbf{D}$ in Eq. (\ref{eq:hoping}) as follows
  \begin{equation}
\phi_\text{Berry}=\oint_\Omega d\mathbf{r}\cdot\mathbf{D}(\mathbf{r})
\end{equation}
where $\Omega$ is the closed loop.

To rotate the Kramers doublet from the local axes ($X$, $Y$ and $Z$) to the global axes ($x$, $y$ and $z$) we use
\begin{equation}
|\tilde{\sigma};\{x\}\rangle=(e^{i\vec{\sigma}\cdot\mathbf{n}_i\theta/2})_{\alpha\sigma}|\alpha;\{X\}\rangle
\end{equation}
where $\{x\}=e^{i\mathbf{l}_i\cdot\mathbf{n}_i\theta}\{X\}$.  The hopping process on the bond $\mathbf{r}_{12}$ is given as
\begin{eqnarray}
\label{eq:soc_Ir}
  t_{12}=\sum_{\sigma\sigma'} \left(t_{\sigma\sigma'}(\mathbf{r}_1,\mathbf{r}_2) c_{1\sigma}^\dag c_{2\sigma'}+\text{h.c.}\right)
\end{eqnarray}
with the hopping parameter
\begin{eqnarray}
  t_{\sigma\sigma'}(\mathbf{r}_1,\mathbf{r}_2)
  &\sim&(e^{-i \vec{\sigma}\cdot\mathbf{n}_{12}\theta})_{\sigma\sigma'}~_1\langle  \pm1|\text{O}_p\rangle \langle\text{O}_p| \pm1\rangle_2\nonumber\\
  &\equiv&-t(e^{-i \vec{\sigma}\cdot\mathbf{n}_{12}\theta})_{\sigma\sigma'}
\end{eqnarray}
This is exactly the same as in Eq. (\ref{eq:hoping}) when $\mathbf{D}_{12}=\mathbf{n}_{12}\theta$. The Berry phase can be very large on the Kagome lattice resulting a very strong spin-orbit coupling effect.

\end{document}